# Adapting Engineering Education to Industrie 4.0 Vision

Selim Coşkun[1], Yaşanur Kayıkcı[1], Eray Gençay*[2]

*Corresponding Author

[1] Turkish-German University, Istanbul, Turkey

[2] University of Tübingen, Tübingen, Germany

**Abstract** Industrie 4.0 is originally a future vision described in the high-tech strategy of the German government that is conceived upon the information and communication technologies like Cyber-Physical Systems, Internet of Things, Physical Internet and Internet of Services to achieve a high degree of flexibility in production, higher productivity rates through real-time monitoring and diagnosis, and a lower wastage rate of material in production. An important part of the tasks in the preparation for Industrie 4.0 is the adaption of the higher education to the requirements of this vision, in particular the engineering education. In this work, we introduce a road map consisting of three pillars describing the changes/enhancements to be conducted in the areas of curriculum development, lab concept, and student club activities. We also report our current application of this road map at the Turkish-German University, Istanbul.

## 1 Introduction

Digitalization is one mega trend of the century and holds the potential to drastically transform various industries and production techniques (Gulati and Soni, 2015). Based on this trend, the term "Industrie 4.0", also known as industry 4.0, has emerged, which is defined as *digitization of the manufacturing sector, with embedded sensors in virtually all product components and manufacturing equipment, ubiquitous cyber-physical systems, and analysis of all relevant data* (McKinsey & Co., 2015). Industrie 4.0 is originally a future vision described in the high-tech strategy of the German government that is conceived upon the information and communication technologies including initiatives such as the *Industrial Internet, Factories of The Future, Internet of Things, Physical Internet, Internet of Services* and *Cyber-Physical Systems*, to achieve a high degree of flexibility in production, higher productivity rates through real-time monitoring and diagnosis and a lower wastage rate of material in production. Cyber-connected manufacturing systems improve efficiency and optimize operations but also have the potential to change the way manufacturers and industrial companies run their business.

Industrie 4.0 took up a pioneering role in industrial IT, which is currently revolutionizing the manufacturing engineering. Many industrialized countries also have already begun with adapting their industrial infrastructure to meet the requirements of the Industrie 4.0 vision. An important part of the tasks in the preparation for Industrie 4.0 is the adaption of the higher education to the requirements of this vision, in particular the engineering education. As Turkish-German University within this changing industry environment, our ultimate aim is to educate outstanding engineers who will contribute and grow with the digitalized world of the future. In this work, we introduce a road map consisting of three pillars describing the changes/enhancements to be conducted in the areas of curriculum development, laboratory concept, and student club activities. We also report our current application of this road map at the Turkish-German University, Istanbul. According to this, first pillar is the implementation of the Industrie 4.0 concept in the *curriculum* of various engineering and science departments, which reveals synergistic benefits of different expertise areas and helps the application and improvement of Industrie 4.0 concept in numerous areas. Second, a *Laboratory - Lego-Lab* is to be realized, where the students work on Industrial Lego Designs using Lego Mindstorms and understand the application of the Industrie 4.0 concept by simulating real production lines. Additionally, Kolb's Experiential Learning Theory is integrated into this part in order to improve student learning experience in Lego-Lab. This theory is designed around four phases: (i) concrete experience, (ii) reflective observation, (iii) abstract conceptualization and (iv) active experimentation. The final and complementary pillar is the establishment of a *student club*, where students work on different aspects of Industrie 4.0. In addition, this club serves as an intermedium for various student and research projects, organization of conferences, and events to introduce and disseminate the Industrie 4.0 vision.

## 2 Methods

In this part, a generic framework of Industrie 4.0 engineering education at Turkish-German University as seen in Figure 1 is presented. The framework consists of three main stages, namely curriculum, laboratory and student club. These pillars are interrelated and even dependent on each other. Furthermore, they are surrounded by the theory of Kolb's Experiential Learning Theory as well as Industrie 4.0 technologies and methods which are conducted along with scientific research including developed ideas and prototypes, running projects at Turkish-German University.

## 2.1 Curriculum

The Industrie 4.0 vision is implemented with a content of curricula into existing courses and new study modules are designed in order to adapt this vision into the engineering education. The module specifications of existing courses are explicitly documented and intersection areas to the Industrie 4.0 vision are determined. Finally, the courses are connected with the practical exercises in the laboratory. Teaching materials for courses with regard to Industrie 4.0 are prepared in order to train the students. Bringing together theoretical and practical units, the curriculum enables students to obtain basic knowledge of Industrie 4.0 relevant themes and experience real business cases in order to provide hands-on exercises for laboratory stage. Primarily, the curricula of all manufacturing related engineering programs are to be adapted in this manner, but also programs like computer science/engineering or material science/engineering should be enhanced with elective courses or study tracks relating to Industrie 4.0.

## 2.2 Laboratory

The practical units are completed with visualization software tools or with simulators at the laboratory. The foundation of a "Visual Production Lab", where computer-aided design and manufacturing (CAD/CAM) with enterprise resource planning (ERP) are to be carried out, Materials and Logistics determined and 3D printed factories generated, will be one of the most important pillars when adapting the engineering education to Industrie 4.0. The aforementioned Lego Design Concept would also be used to optimize factory design, since it enables dynamic simulation of different production entities with moving parts and supports the static 3D printed factory model.

Lego Mindstorms systems provide programmable brick computers, modular motors and sensors, and a variety of Lego Technic elements, which can be used to simulate real production lines. Sensors like touch, light, distance, sound, and servo motor/rotation sensors and programmable brick computers provide the components needed for the development of intelligent manufacturing models and concepts that are central to the Industrie 4.0 vision. With the Lego factory to be established within the scope of this study, sample production lines for different products will be set up and alternative production plans will be designed. according to the Industrie 4.0 concept.

A drone will be used to investigate a real-life landscape and capture the overlapping images while in motion, images are converted to georeferenced (i.e. geographical information system-GIS) 2D data maps and 3D models via using drone-based mapping software for the foundation of a real factory. This drone-based factory inspection helps measure the stockpiles, efficiency of factory layout, performance of assembly lines, etc. As a result, all the simulation data will be compared to real data and the obtained feedback will be used to consummate the whole design. For this purpose, a 3D Printing Lab in the Visual Production Lab is to be established. Here, realistic models of digitally designed factory will repeatedly be created and the obtained feedback will be used to remedy the flaws of the design. In addition to that, production of non-standard Lego parts will be carried out and accordingly mass production scenarios specific to different demands will be realized. Moreover, a research on material properties used in 3D printing (and also 3D pen) can also be carried out. The main aim in material research should be to develop a strong, durable and low melting temperature polymer, which as a result might lead to rapid manufacturing beside rapid prototyping (RP). The rapid prototyping refers to a process used in various industries for rapidly creating a representative system or part before final release or commercialization.

Additive manufacturing (AM) is the formalized term for what used to be called rapid prototyping and what is nowadays popularly called 3D Printing. In mass customization, one of the key concepts is the capability of flexible production, i.e. production of individualized products on the same line subsequently, at the costs near that of mass production. In order to achieve that, additive manufacturing has been proposed as a flexible production model. So, the developments in 3D printing technology might lead the transformation from rapid prototyping to rapid manufacturing and ultimately to ideal mass customization. As a result, the synergetic efforts of the students and also researchers in Visual Production Lab and Lego-Lab will make a significant contribution for the adaption of the engineering education to Industrie 4.0, since an ideal engineering education should contain the combination of scientific research and industrial application.

### 3.1.1 Implementing Kolb's theory to laboratory

Experiential and simulation learning techniques are used in a wide range of fields from quality (Wang, 2004) to product development (Holmqvist, 2004), process re-engineering (Smeds and Riis, 1998) and supply chain management (Carbonneau et al., 2008), as contributors in order to create a learning and practicing environment that maximizes learners' skills in learning from their own experience, the full potential for learning can be realized, moreover the costs during learning process in case of failures can be avoided. Basically, learning comes from three principal sources: learning from content, learning from experience and learning from feedback (Kolb, 1984; Kirby, 1992):

1. *learning from content:* the discovery of new ideas, principles and concepts;
2. *learning from experience:* an opportunity to apply content in an experiential environment;
3. *learning from feedback:* the results of actions taken and the relationship between the actions in the experiment and performance

The experiential learning method creates an environment that requires the participant to be involved in some type of personally meaningful activity. Such an environment allows the participant to apply prior knowledge of theory and principles while developing commitment to the exercise and experiencing a real sense of personal accomplishment or failure for the results obtained (Keys and Wolfe, 1990). In order to bring change in behavior, attitudes and knowledge, a circular four-stage experiential learning cycle model developed by Kolb (Kolb, 1984) is used. Kolb's Experiential Learning Theory depicted in Figure 2 puts emphasis on sensory and emotional engagement in the learning activity. Effective learning is seen when a person progresses through a cycle of four stages: of (i) having a *concrete experience* followed by (ii) *observation* of and *reflection* on that experience which leads to (iii) the formation of *abstract concepts* (analysis) and generalizations (conclusions) which are then (iv) used to test hypothesis in future situations, resulting in new *experiences*.

At Turkish-German University the learning process at the laboratory in the frame of Industrie 4.0 engineering education is provided by implementation of Kolb's Experiential Learning Theory. Engineering students are divided into the groups at the laboratory and each group needs to get a certain task from the instructor about Industrie 4.0, for example, designing an assembly line of digital car factory. The steps are listed as follows:

(i) *concrete experience:* which means direct practical experience by performing a new task. In our activities, concrete experience corresponds to a set of step-by-step instructions demonstrating a new concept. Originally, all activities in Lego-Lab are designed based on this approach. In the illustrative example, students follow step-by-step instructions to learn and get a broad understanding of Industrie 4.0 and its applications such as assembly line planning with mixed-model scenarios. The activity instructions are written for novice users and are very descriptive so that the students can complete the activity even though they have in previous experience in the field.

(ii) *reflective observation:* which includes activities such discussion and reflective questions that require students to reflect on their hands-on experiences (hands-on exercises) that enable students to work in Industrie 4.0-conform environments. Our strategy is to divide an activity into smaller sections and include reflective activities for each section. This strategy also helps the instructor phase the activity across multiple groups. In the illustrative example, after completing the first section, students are asked to analyze components of assembly line and discuss questions such as why and what kind of robotics they have to put into the assembly line. Reflective observation activities should foster student-to-student interaction in order to achieve a higher level of reflection. Group work is a particularly effective strategy to promote meaningful reflection in short classroom activities. For example, instead of asking students to analyze their own response individually, asking them to compare the responses with group members and list similarities and differences may lead to a higher level of reflection.

(iii) *abstract conceptualization:* through abstract conceptualization, students are expected to create a theoretical model and generalization of what was performed. Generally, this stage could be difficult to achieve in short hands-on activities. Class or peer-to-peer discussions are helpful to connect the learning experience to the overall theory. At this stage, instructor intervention is important. In the illustrative example, students are asked to create a digitalization scenario by applying robotics technologies based on the steps that they perform. In this part, also brainstorming applications such as mind-mapping software, etc. are used. After this question, a class discussion led by the instructor may help students solidify a mental picture of robot configurations at the assembly line. Another useful strategy is the utilization of generalization questions. For example, in the illustrative example students are asked to compare what they performed in an earlier activity about robotics (i.e. kinematics) and to list the advantages and disadvantages of a type of robot (i.e. articulated robot, parallel arm robot, SCARA robot etc.). Such generalization questions can be combined with the next stage of active experimentation.

(iv) *active experimentation:* At this stage, the student is ready to plan and try out another concrete experience. We use two strategies in this stage. The first strategy is to give students a new task, albeit similar to what was performed in the concrete experience stage, but without providing step-by-step instructions. For example, students are asked to configure a SCARA robot by applying some program comments. By this time, they should be able to achieve this task without detailed instructions. The second strategy is to combine a few related topics in the same activity such that the later topics build on the former ones.

## *3.3 Student Club*

A student club will enable the students to study the different aspects of Industrie 4.0. It is considered that the activities of the club should be conducted under the roof of an Open Innovation Office or research lab, which will operate in parallel with the student club and focus on scientific developments on the topic "Industrie 4.0". The objectives of the student club will include realization of student and research projects, organization of conferences, and events to introduce and disseminate the Industrie 4.0 concepts. One of the first projects that will be carried out by the club is considered to be the application of the Industrie 4.0 scenarios to the production models at a Lego-Lab. The Lego-Lab will be so designed that the students can work on Industrial Lego Designs using Lego Mindstorms and understand the application of the Industrie 4.0 concept by simulating real production lines.

An example application on a Lego factory model would be a scenario, where a car product variant is produced individually according to its order code, which is read by robots as it proceeds through the production line. Order codes can be encoded as color sequences that can be recognized by a standard Lego color sensor. Alternatively, a barcode/QR-code scanner could be used to read order codes. The data coming from the sensor will be then processed by a Lego intelligent brick coordinating the individual robot actions. At the end, the production line should function as fast as in the batch production but produce completely individualized cars. Similar to this example, different scenarios can be realized or simulated using the Lego Design Concept. The fact that this approach is very user-friendly and easy to understand makes it an ideal asset for students to embrace and benefit from.

The Lego-Lab will provide a production line consisting of multiple modules. Lego computer bricks that control each robot on the production line are programmable and support i.a. C/C++, Java, Phyton and Visual Basic. The following can be realized as first projects by the student club:

1. Integration of the Lego production modules with a cloud server in order to be monitored and managed through this server (Schlechtendahl et al., 2014). In the project, data from the active sensors on the Lego production modules are transferred to the cloud server and then stored in suitable data structures for further analysis. As database technology, a No-SQL database should be used, since sensor data can accumulate very fast and high in amount, which makes a relational approach unfeasible considering the query performance.

2. Printing new Lego parts using a 3D printer to enable the variant rich production of a certain product. As the first product example, transport vans can be chosen. These vehicles are especially suitable for a lab application in Industrie 4.0, since they provide a very high number of variants.

3. Transformation of Lego production modules from mass production model (Industrie 3.0) to individualized mass production model (Industrie 4.0). At the end of this project, the code on a chassis that enters the production line will be scanned by a sensor and the production order belonging to this code will be queried on the cloud server. This inquiry will be processed for each of the modules in the production line as the chassis check in. According to the response from the server, robot will decide immediately from which feeder it should take parts to continue with the montage. Thus, the production line will be working on a speed comparable to the mass production but allowing the line to put out a new vehicle configuration each time.

4. Design of an optimal production line according to the product tree. Again here, transport vans can be chosen as a variant rich product. Students will analyze this product starting from the chassis taking in account different variants that can arise in the montage process using a product tree. According to the results of this analysis, they will redesign the fabric layout by rearranging modules, robots, stations and feeders optimally. The resulting new layout design will be applied on the Lego modules and production time results from the new and the old layouts will be compared.

5. Addition of new Lego modules. Students will enhance the Lego production line with new Lego Mindstorm parts that will be procured and so will enable the production of new product examples.

# 4 Conclusions

In this work, we proposed a generic framework for Industrie 4.0 education that consisted of curriculum, laboratory, and student club components to adapt the engineering education to Industrie 4.0 vision. In the curriculum component, we determined new study modules to be introduced, and changes to the existing study modules. In the next step, we designed two main labs to address the changes we made on the curricula, namely the Visual Production Lab, and the Lego-Lab. While defining the relation between the course hours and hands-on laboratory units, and how these laboratory units are to be conducted, we used Kolb's Experiential Learning Theory. In the last step, we showed how a student club can complement the changes defined in the first two steps. Students can take initiative for Industrie 4.0 related projects under the roof of a student club, which in turn supports the implementation of the active experimentation stage of the Kolb's Experiential Learning Theory. The preliminary results from our implementation of this framework at the Turkish-German University showed that it was feasible to apply such a framework and the adopted underlying theory of Kolb to adapt the engineering education to Industrie 4.0 vision.

**Acknowledgments** This study was supported by Turkish-German University Scientific Research Projects Commission under the grant no: 2016BM0015.

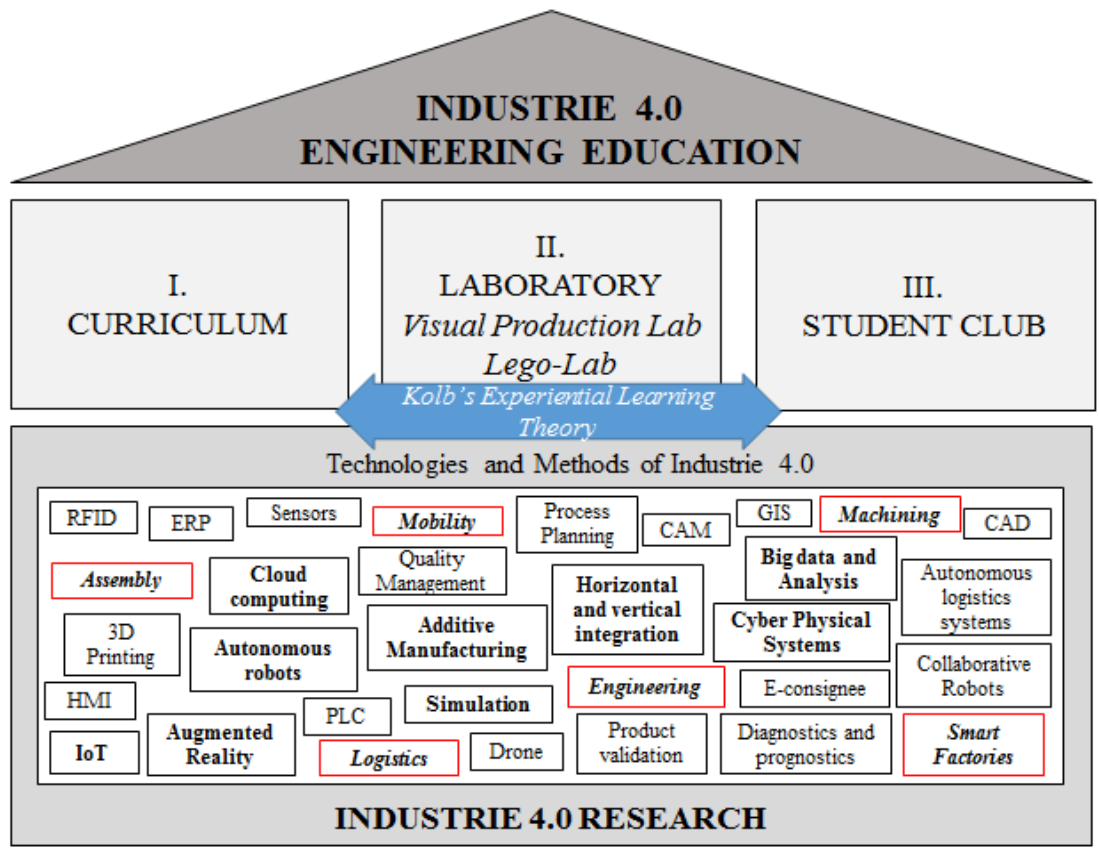

**Fig. 1.** Generic framework of Industrie 4.0 engineering education

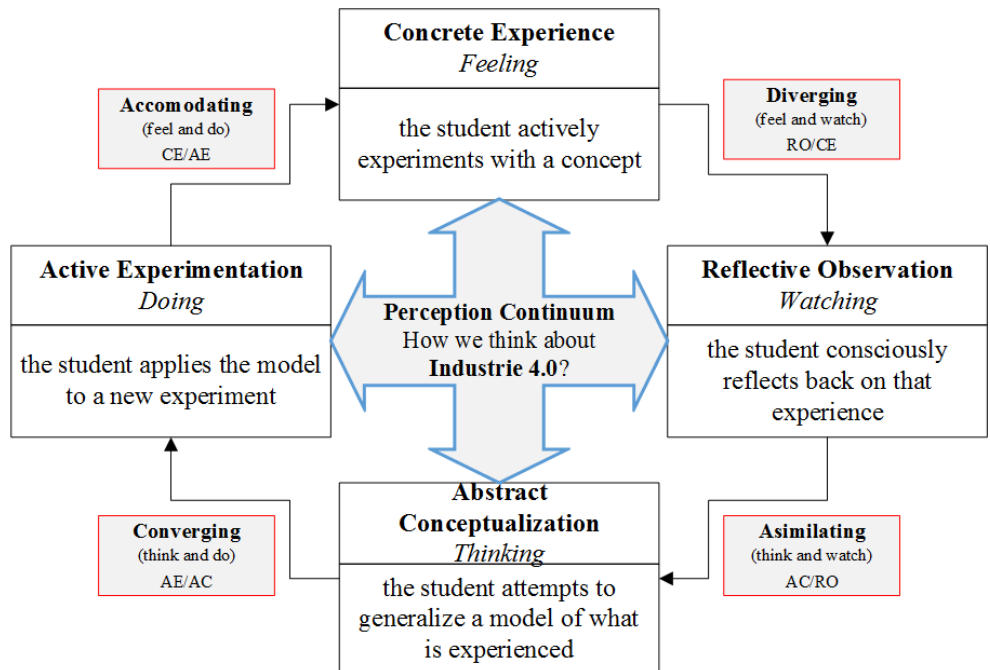

**Fig. 2.** Kolb's learning styles model and experiential learning theory